\documentstyle[sprocl,axodraw]{article}

\makeatletter

\def\nl{\nonumber\\}

\def\beq{\begin{equation}}
\def\eeq{\end{equation}}
\def\beqar{\begin{eqnarray}}
\def\eeqar{\end{eqnarray}}

\def\de{\delta}
\def\la{\lambda}
\def\si{\sigma}
\def\refeq#1{\mbox{(\ref{#1})}}
\def\mathswitchr#1{\relax\ifmmode{\mathrm{#1}}\else$\mathrm{#1}$\fi}
\newcommand{\PW}{\mathswitchr W}
\newcommand{\PZ}{\mathswitchr Z}
\def\mathswitch#1{\relax\ifmmode#1\else$#1$\fi}
\newcommand{\MW}{\mathswitch {M_\PW}}
\newcommand{\MZ}{\mathswitch {M_\PZ}}

\newcommand{\muD}{{\mu_0}}
\newcommand{\betacoeff}[1]{b_{#1}}
\newcommand{\ctwo}{C_{\si}}

\newcommand{\etal}{{\it et al.}}
\newcommand{\NLLA}{\stackrel{\mathrm{NLL}}{=}}
\newcommand{\SUtwo}{\mathrm{SU}(2)}
\newcommand{\Uone}{\mathrm{U}(1)}
\newcommand{\ri}{\mathrm{i}}
\newcommand{\veps}{{\varepsilon}}
\newcommand{\elm}{{\mathrm{em}}}
\newcommand{\QED}{{\mathrm{QED}}}
\newcommand{\Epsinv}[1]{\veps^{- #1}}
\newcommand{\Eps}[1]{\veps^{#1}}
\newcommand{\univfact}[2]{I(#1,#2)}
\newcommand{\NLLfact}[3]{J(#1,#2,#3)}
\newcommand{\FF}[2]{F_{#1}^{(#2)}}
\newcommand{\gb}{g_1}
\newcommand{\gw}{g_2}
\newcommand{\normfact}{N_{\veps}}

\begin{document}

%%%%%%%%%%%%%%%%%%%%%%%%%%%%%%%%%%%%
%  TITLEPAGE
%%%%%%%%%%%%%%%%%%%%%%%%%%%%%%%%%%%%
{
\thispagestyle{empty}
\def\thefootnote{\fnsymbol{footnote}}
\setcounter{footnote}{1}
\null
%\draftdate
\hfill   TTP04-21\\
\strut\hfill  SFB/CPP-04-42\\
\strut\hfill hep-ph/0409091
\vskip 0cm
\vfill
\begin{center}
{\Large \bf
Next-to-leading electroweak logarithms at two loops
\par}
 \vskip 1em
{\large
{\sc S.~Pozzorini\footnote{pozzorin@particle.uni-karlsruhe.de} }}
\\[.5cm]
{\it Institut f\"ur Theoretische Teilchenphysik, 
Universit\"at Karlsruhe \\
D-76128 Karlsruhe, Germany}
\par
\end{center}\par
\vskip 1.0cm 
\vfill 
{\bf Abstract:} \par 
We review a recent calculation of the two-loop 
next-to-leading logarithmic electroweak corrections to the form factors for  
massless chiral fermions coupling to an $\SUtwo\times \Uone$ singlet gauge boson.

% abstract in text format
%We review a recent calculation of the two-loop next-to-leading logarithmic electroweak corrections to the form factors for massless chiral fermions coupling to an SU(2)xU(1) singlet gauge boson.

\par
\vskip 1cm
\noindent
September 2004 
\par
\null
\setcounter{page}{0}
}
\clearpage
\setcounter{footnote}{0}

%%%%%%%%%%%%%%%%%%%%%%%%%%%%%%%%%%%%%%%%%%%%%%%%%%
%                                                %
%    BEGINNING OF TEXT                           %
%                                                %
%%%%%%%%%%%%%%%%%%%%%%%%%%%%%%%%%%%%%%%%%%%%%%%%%%

\title{NEXT-TO-LEADING ELECTROWEAK LOGARITHMS AT TWO LOOPS}

\author{STEFANO POZZORINI}

\address{Institut f\"ur Theoretische Teilchenphysik, Universit\"at Karlsruhe, D-76128 Karlsruhe}
\par

\maketitle\abstracts{
We review a recent calculation of the two-loop 
next-to-leading logarithmic electroweak corrections 
to the form factors for  
massless chiral fermions coupling to an $\SUtwo\times \Uone$ singlet 
gauge boson.
}

\section{Introduction}
\label{se:intro}%\refse{intro}
In the energy region $\sqrt{s}\gg \MW\sim\MZ$,
the electroweak (EW) radiative corrections are strongly enhanced by logarithms of the form%
\footnote{The terms with $j=0,1,2,\dots$ are denoted as leading logarithms (LLs),
next-to-leading logarithms (NLLs), next-to-next-to-leading logarithms (NNLLs), 
etc.}
$\alpha^l\log^{2l-j}{\left({s}/{\MW^2}\right)}$, with  $j=0,1,\dots,2l$.
At $\sqrt{s}\sim 1$ TeV, these logarithms
yield one-loop corrections of tens of per cent
and two-loop corrections of several per cent.
These EW corrections will be important for the interpretation of the measurements at the Linear Collider.

At one loop, the EW LLs and NLLs 
are now well understood
\cite{Denner:2001jv,Pozzorini:rs}. 
Resummation prescriptions for the LL\cite{Fadin:2000bq}
and NLL\cite{Kuhn:2000nn,Melles:2001gw} corrections to arbitrary processes exist and the EW logarithmic corrections to 4-fermion processes have been resummed 
up to the NNLLs\cite{Kuhn:2001hz}.
At the TeV scale, the leading and subleading logarithms have similar 
size\cite{Kuhn:2000nn,Kuhn:2001hz} and these latter must be under control in order to reduce the theoretical error at the few per mille level. The above resummations were obtained by means of the infrared evolution equation\cite{Fadin:2000bq} (IREE) assuming exact SU(2)$\times$U(1) and  $\Uone_\elm$ gauge symmetry for the regimes above and below the EW scale, respectively.
This approach relies on the assumption that several aspects of symmetry breaking can be neglected in the high-energy limit. In particular, the following aspects are neglected.
(i) The couplings proportional to the vacuum expectation value (vev);
(ii) The weak-boson mass gap ($\MW=\MZ$ approximation);
(iii) The mixing between neutral gauge bosons.

These resummations need to be checked by means of two-loop calculations based on the electroweak Lagrangian, where all effects from symmetry breaking are taken into account.
It was alredy confirmed that the LL corrections\cite{Melles:2000ed,Hori:2000tm,Beenakker:2000kb} as well as the angular-dependent subset of the NLL corrections\cite{Denner:2003wi} to arbitrary processes exponentiate 
as predicted by the IREE.
At the NLL level (and beyond) only few two-loop calculations exist\cite{Feucht:2003yx,Pozzorini:2004rm,Feucht:2004rp}.
Here we review a calculation of the one- and two-loop virtual EW NLL corrections to the form factors for massless chiral fermions coupling to an $\SUtwo\times \Uone$ singlet gauge boson\cite{Pozzorini:2004rm}.

\section{Perturbative and asymptotic expansion}
\label{se:definitions}%\refse{se:definitions}

%%%%%%%%%%% Feynman diagram  %%%%%%%%%%%%%%%%%%
\newcommand{\diaggeneric}{
\begin{picture}(120.,104.)(-28.,-52.)
\Gluon(-53.6656,0.)(0.,0.){3.2}{5}
\ArrowLine(0.,0.)(24.,12.)
\ArrowLine(24.,12.)(48.,24.)
\ArrowLine(24.,-12.)(0.,0.)
\ArrowLine(48.,-24.)(24.,-12.)
\GCirc(4.47214,0.){17}{0.8}
\Text(27.668,14.6453)[br]{}
\Text(28.,-14.)[tr]{}
\end{picture}
}

%%%%%%%%%%%%%%%%%%%%%%%%%%%%%%%%%%%%%%%%%%%%%%

The vertex function for 
an SU(2)$\times$U(1) singlet gauge boson (which might be for instance a gluon) 
coupled to massless fermions can be written in terms of 
chiral form factors $F_\pm$ as
\vspace{-10mm}
\beqar\label{diagramgeneric}%\refeq{diagramgeneric}
\vcenter{\hbox{\diaggeneric}}
\hspace{-13mm}=
\ri 
\mbox{$\bar{u}(p_1) \gamma^\nu \biggl(\omega_- F_-+\omega_+ F_+\biggr) v(p_2)$}
\eeqar
\\[-8mm]
with $\omega_\pm=\frac{1}{2}(1\pm\gamma^5)$ and the 
perturbative expansion 
\beqar\label{pertserie1}%\refeq{pertserie1}
F_\si=
\FF{\si}{0}
\left[
1+\sum_{l=1}^\infty
\left(\frac{\alpha}{4\pi}\right)^l
\normfact^l
\de \FF{\si}{l}
\right]
\quad\mbox{with}\quad
\normfact=
\frac{1}{\Gamma(1-\veps)}
\left(\frac{4\pi\muD^2}{-s}\right)^{\veps}
\eeqar
in  $D=4-2\veps$ dimensions.
The one- and two-loop EW corrections, 
$\de \FF{\si}{1}$
and $\de \FF{\si}{2}$,
were evaluated  in the asymptotic region 
$s=(p_1+p_2)^2 \gg \MW^2\sim\MZ^2$,
including mass singular logarithms 
$L=\log\left({-s}/{\MW^2}\right)$ 
originating from massive weak bosons 
and $1/\varepsilon$ poles
from massless photons.
Corrections of order $\MW^2/s$ were neglected.
The relevant Feynman diagrams were computed 
to NLL accuracy using an algorithm\cite{Denner:2004iz}  based on sector decomposition\cite{Hepp:1966eg,Roth:1996pd,Binoth:2000ps}.
The NLL approximation includes all terms of the order
$\alpha^l\veps^{k}\log^{j+k}(s/\MW^2)$ with 
$j=2l,2l-1$. 
Contributions depending on $\MZ/\MW$ were also included.

\section{Results}
\label{se:oneloop}%\refse{se:oneloop}
The one-loop corrections read%
\footnote{The corrections are expressed in terms of  the hypercharge $Y_\si$, 
the weak isospin $T^3_\si$ and the SU(2) Casimir operator $C_\si$ 
for left-handed ($\si=-$) and  right-handed ($\si=+$) fermions.
The electromagnetic charge of the fermions is denoted as $Q$ whereas  $g_1$ and $g_2$ represent the U(1) and SU(2) coupling constants, respectively.
}
\beqar\label{oneloopren}%\refeq{oneloopren}
\lefteqn{
e^2 \de \FF{\si}{1}\NLLA
\left[
\gb^2\left(\frac{Y_\si}{2}\right)^2
+\gw^2\ctwo
\right]
\univfact{\veps}{\MW}
}\quad&&
\nl&&{}+
\left[
\gb^2\left(\frac{Y_\si}{2}\right)^2 
+\gw^2(T_\si^3)^2-e^2 Q^2
\right]
\Delta
\univfact{\veps}{\MZ}
+
e^2 Q^2
\Delta\univfact{\veps}{0},
\eeqar
where%
\footnote{
Note that the one-loop corrections are expanded up to order
$\veps^2$. These higher-order terms in the $\veps$-expansion
must be taken into account in the relation \refeq{ewresummed1}
between one- and two-loop 
mass singularities.}
\beqar\label{Ifunc}%\refeq{Ifunc}
\univfact{\veps}{\MW}&\NLLA&
-L^2
-\frac{2}{3}L^3\Eps{}
-\frac{1}{4}L^4\Eps{2}
+3L
+\frac{3}{2}L^2\Eps{}
+\frac{1}{2}L^3\Eps{2}
+{\mathcal{O}}(\varepsilon^3)
,\nl
\univfact{\veps}{\MZ}&\NLLA&
\univfact{\veps}{\MW}+
\log\left(\frac{\MZ^2}{\MW^2}\right)\left(
2L
+2L^2\Eps{}
+L^3\Eps{2}
\right)
+{\mathcal{O}}(\varepsilon^3)
,\nl
\univfact{\veps}{0}&\NLLA&
-2\Epsinv{2}
-3\Epsinv{1}
,
\eeqar
and $\Delta\univfact{\veps}{m}=\univfact{\veps}{m}-\univfact{\veps}{\MW}$.
The corrections are split into a contribution
$\univfact{\veps}{\MW}$, which corresponds to a symmetric 
SU(2)$\times$U(1) theory with all gauge boson masses equal to $\MW$,
and additional contributions $\Delta \univfact{\veps}{\MZ}$ and
$\Delta \univfact{\veps}{0}$, which  result from the Z--W and 
$\gamma$--W mass gaps.
It was found that the two-loop corrections can be written 
in terms of the one-loop functions \refeq{Ifunc}
in a form that is strictly analogous to 
Catani's formula for two-loop mass singularities in massless QCD\cite{Catani:1998bh}. In the on-shell renormalisation scheme with the 
electromagnetic coupling constant renormalised at the scale $\MW$
we have
\beqar\label{ewresummed1}%\refeq{ewresummed1}
e^4 \de \FF{\si}{2}
&\NLLA&
\frac{1}{2}
\left[e^2 \de \FF{\si}{1}
\right]^2
+e^2 \left[
\gb^2\betacoeff{1}^{(1)}
\left(\frac{Y_\si}{2}\right)^2
+\gw^2\betacoeff{2}^{(1)}
\ctwo
\right]
\NLLfact{\veps}{\MW}{\MW}
\nl&&{}
+
e^4 \betacoeff{\QED}^{(1)}
Q^2
\Delta \NLLfact{\veps}{0}{\MW}
.
\eeqar
The first term on the right-hand side
can be regarded as the result of
the exponentiation of the  one-loop corrections.
The additional contributions, which are associated to the one-loop 
$\beta$-function coefficients\cite{Pozzorini:2004rm}
$\betacoeff{1}^{(1)}$,
$\betacoeff{2}^{(1)}$
and $\betacoeff{\QED}^{(1)}$,
are proportional to the functions $J$ and  $\Delta J$,
defined as 
\beqar\label{NLLfacts}%\refeq{NLLfacts}
\NLLfact{\veps}{m}{\mu}=
\frac{1}{\veps}\left[
\univfact{2\veps}{m}
-\left(\frac{-s}{\mu^2}\right)^\veps
\univfact{\veps}{m}
\right]
\eeqar
and $\Delta \NLLfact{\veps}{m}{\mu}=
\NLLfact{\veps}{m}{\mu}
-
\NLLfact{\veps}{\MW}{\mu}$.
The contributions proportional to
$\left[\univfact{\veps}{\MW}\right]^2$ 
%${\univfact{\veps}{\MW}}^2$ 
and $\NLLfact{\veps}{\MW}{\MW}$ correspond to the two-loop corrections within an unbroken $\SUtwo\times\Uone$  theory where all gauge bosons have mass $\MW$.
The only  two-loop  terms depending on the Z--W mass gap
are the NLLs proportional to  $\log(\MZ^2/\MW^2)$ that 
arise from the squared 
one-loop corrections in \refeq{ewresummed1}
via the term
$\Delta\univfact{\veps}{\MZ}$ in \refeq{oneloopren}.
This means that such Z--W mass-gap terms
exponentiate.
The infrared divergent $1/\varepsilon$ poles are isolated in the terms proportional to
$Q^2\Delta\univfact{\veps}{0}$ and 
$Q^2\betacoeff{\QED}^{(1)}\Delta\NLLfact{\veps}{0}{\MW}$.
Such terms originate from the $\gamma$--W mass gap and
correspond to QED corrections with photon mass $\la=0$ 
subtracted at $\la=\MW$.
Apart from the NLLs proportional to
$\log(\MZ^2/\MW^2)$, 
the result \refeq{ewresummed1} is independent on 
the vev, the Z--W mass gap
and the weak mixing angle.
This simple behaviour of the two-loop form factors results from 
cancellations between NLLs from different 
diagrams and is ensured by the relations between the vev,
the weak-boson masses 
and the weak mixing angle\cite{Pozzorini:2004rm}.
This result confirms the assumptions (i)--(iii) discussed in 
the introduction and the IREE-approach\cite{Fadin:2000bq,Kuhn:2000nn,Melles:2001gw} 
to resum the NLL EW  corrections.

\end{document}